\documentclass{article}
\usepackage{spconf,amsmath,graphicx}
\usepackage[printonlyused, nolist]{acronym}
\usepackage{color}
\usepackage{amssymb, mathtools}
\usepackage{booktabs,bm}
\usepackage{array,multirow}
\usepackage{tikz, pgfplots} 

\definecolor{LMSred}{rgb}{0.80,0.20,0.20} 
\definecolor{LMSgray}{rgb}{0.60,0.60,0.60}
\definecolor{LMSCyan}{rgb}{0.60,1,1}
\definecolor{LMSMagenta}{rgb}{0.965,0.294,1}
\definecolor{LMSYellow}{rgb}{0.95,0.95,0.0}
\definecolor{LMSOrange}{rgb}{1,0.5,0}
\definecolor{LMSlightblue}{rgb}{0.5,0.8,1}
\definecolor{LMSlightred}{rgb}{1,0.294,0.294}
\definecolor{LMSblue}{rgb}{0.216,0.255,1}
\definecolor{LMSgreen}{rgb}{0.15,0.7,0.15}
\definecolor{LMSlightgreen}{rgb}{0.35,0.90,0.35}
\definecolor{LMSdarkgreen}{rgb}{0.01,0.5,0.01}

\newcommand{\WK}[1]{#1}
\newcommand{\WKtwo}[1]{#1}

\newcommand{\SER}{\mathrm{SER}}
\newcommand{\ie}{, i.e., }
\newcommand{\eg}{, e.g., }
\newcommand{\wrt}{w.r.t. }

\newcommand{\RTF}{h}
\newcommand{\RTFest}{\hat{\RTF}}
\newcommand{\ATF}{a}
\newcommand{\RTFVec}{\mathbf{\RTF}}

\newcommand{\RTFRI}{\RTFVec}
\newcommand{\RTFRIrec}{\tilde{\RTFVec}}
\newcommand{\RTFRIest}{\hat{\RTFVec}}

\newcommand{\MicSig}{x}
\newcommand{\SrcSig}{s}
\newcommand{\SpatialImage}{c}
\newcommand{\NoiseSig}{n}

\newcommand{\MicSigVecBB}{\mathbf{\MicSig}}

\newcommand{\MicSigSet}{\mathbb{X}}

\newcommand{\Embedding}{z}
\newcommand{\EmbeddingVec}{\mathbf{\Embedding}}

\newcommand{\Mean}{\mu}

\newcommand{\MeanVec}{\bm{\Mean}}
\newcommand{\VarianceVec}{\WK{\bm{v}}}
\newcommand{\CovMat}{\bm{\Sigma}}
\newcommand{\VarianceDec}{\sigma^2}

\newcommand{\EncoderParams}{\bm{\phi}}
\newcommand{\DecoderParams}{\bm{\theta}}

\newcommand{\Encoder}{\mathtt{enc}_{\EncoderParams}}
\newcommand{\Decoder}{\mathtt{dec}_{\DecoderParams}}

\newcommand{\EncoderModel}{q_{\EncoderParams}}
\newcommand{\DecoderModel}{p_{\DecoderParams}}

\newcommand{\ELBO}{\mathcal{L}_{\mathrm{VAE}}}
\newcommand{\CostFunVAE}{\mathcal{J}_{\mathrm{VAE}}}
\newcommand{\CostFunLS}{\mathcal{J}_{\mathrm{LS}}}

\newcommand{\AutoPSD}{\WK{S}_{11,\FreqIdx\FrameIdx}}
\newcommand{\CrossPSD}{\WK{S}_{12,\FreqIdx\FrameIdx}}

\newcommand{\Gaussian}{\mathcal{N}}

\newcommand{\FreqIdx}{f}
\newcommand{\FreqIdxMax}{F}
\newcommand{\FrameIdx}{t}
\newcommand{\FrameIdxMax}{T}
\newcommand{\MicIdx}{m}
\newcommand{\MicIdxMax}{2}

\newcommand{\BatchIdx}{b}
\newcommand{\BatchSize}{B}

\newcommand{\InputSize}{D}

\newcommand{\BottleneckDim}{d}

\newcommand{\DataIdx}{n}
\newcommand{\DataIdxMax}{N}
\newcommand{\transp}{^\mathtt{T}}
\newcommand{\Avg}{\hat{\mathcal{E}}}
\newcommand{\Norm}[1]{\left\Vert #1 \right\Vert_2}
\newcommand{\Real}[1]{\mathrm{Re}\left\lbrace #1 \right\rbrace}
\newcommand{\Imag}[1]{\mathrm{Im}\left\lbrace #1 \right\rbrace}
\newcommand{\trace}{\mathrm{tr}}

\newcommand{\Expect}{\mathcal{E}}
\newcommand{\KLD}{\mathcal{KL}}

\newcommand{\StepSizeLS}{\alpha}

\title{Manifold Learning-supported Estimation of Relative Transfer Functions for Spatial Filtering}
\name{Andreas Brendel, Johannes Zeitler, and Walter Kellermann\thanks{This work was funded by the Deutsche Forschungsgemeinschaft	(DFG, German Research Foundation) -- 282835863 -- within the Research Unit FOR2457 “Acoustic Sensor Networks”.}}
\address{\fontsize{11}{12}\selectfont\textit{Multimedia Communications and Signal Processing,}
	\textit{Friedrich-Alexander-Universit\"at Erlangen-N\"urnberg,}\\
	\fontsize{11}{12}\selectfont Cauerstr. 7, D-91058 Erlangen, Germany,
	e-mail: \texttt{Andreas.Brendel@FAU.de}}
\begin{document}
\begin{acronym}
	\acro{STFT}{Short-Time Fourier Transform}
	\acro{PSD}{Power Spectral Density}
	\acro{PDF}{Probability Density Function}
	\acro{RIR}{Room Impulse Response}
	\acro{ATF}{Acoustic Transfer Function}
	\acro{FIR}{Finite Impulse Response}
	\acro{FFT}{Fast Fourier Transform}
	\acro{DFT}{Discrete Fourier Transform}
	\acro{MAP}{Maximum A Posteriori}
	\acro{RTF}{Relative Transfer Function}
	\acro{DOA}{Direction of Arrival}
	\acro{SNR}{Signal-to-Noise Ratio}
	\acro{SIR}{Signal-to-Interference Ratio}
	\acro{SDR}{Signal-to-Distortion Ratio}
	\acro{SAR}{Signal-to-Artefact Ratio}
	\acro{LS}{Least Squares}
	\acro{AE}{Autoencoder}
	\acro{VAE}{Variational Autoencoder}
	\acro{BSS}{Blind Source Separation}
	\acro{ELBO}{Evidence Lower Bound Objective}
	\acro{DNN}{Deep Neural Network}
	\acro{KLD}{Kullback-Leibler Divergence}
	\acro{SER}{Signal-to-Error Ratio}
	\acro{OOGP}{Out-Of-Grid Position}
	\acro{WGN}{White Gaussian Noise}
	\acro{AWGN}{Additive White Gaussian Noise}
	\acro{PS}{Point Source}
	\acro{GT}{Ground Truth}
\end{acronym}
\ninept
\maketitle
\begin{abstract}
	Many spatial filtering algorithms used for \WK{voice capture in}\eg teleconferencing applications, can benefit from or even rely on knowledge of Relative Transfer Functions (RTFs). \WK{\WKtwo{Accordingly}, many} RTF estimators have been proposed which, however, suffer \WK{from performance degradation} under acoustically adverse conditions or need prior knowledge on the properties of the interfering sources. \WK{While state-of-the-art RTF estimators ignore prior knowledge about the acoustic enclosure}, audio signal processing algorithms for teleconferencing \WKtwo{equipment are often operating in} the same or at least a similar acoustic enclosure\eg a car or an office, such that training data can be collected. In this contribution, we use such data to train Variational Autoencoders (VAEs) in an unsupervised manner and apply the trained VAEs to enhance imprecise RTF estimates. Furthermore, a hybrid between classic RTF estimation and the trained VAE is investigated. Comprehensive experiments with real-world data confirm the efficacy \WKtwo{for} the proposed method. 
\end{abstract}
%
%%%%%%%%%%%%%%%%%%%%%%%%%%%%%%%%%%%%%%%%%%%%%%%%%%%%%%%%%
\section{Introduction and Signal Model}
\label{sec:intro}
%%%%%%%%%%%%%%%%%%%%%%%%%%%%%%%%%%%%%%%%%%%%%%%%%%%%%%%%%
%
%Teleconferencing connects people all over the globe and bridges long distances by enabling cooperation, social presence and close-to-natural communication. Hence, 
\WK{T}eleconferencing is an essential technology for modern work places that recently became even more important, e.g., due to the increasing number of people working from home during pandemic situations. An integral aspect of such systems is the enhancement of the voices of the conversation partners under adverse acoustic conditions, i.e., the suppression of undesired sources like people talking in the background. As most of the commonly used teleconferencing devices are equipped with multiple microphones, spatial filtering \WK{is an obvious algorithmic choice of maximizing user comfort}. 

In this paper we consider a source signal observed by a pair of microphones in a reverberant and noisy environment. The \ac{STFT}-domain microphone observation $\MicSig_{\MicIdx,\FreqIdx\FrameIdx}$ at frequency index $\FreqIdx\in\{1,\dots,\FreqIdxMax\}$, time frame index $\FrameIdx\in\{1,\dots,\FrameIdxMax\}$ and microphone index $\MicIdx\in\{1,2\}$ can be modeled by
\begin{equation}
	\WK{\MicSig_{\MicIdx,\FreqIdx\FrameIdx} = \ATF_{\MicIdx,\FreqIdx}\SrcSig_{\FreqIdx\FrameIdx} + \NoiseSig_{\MicIdx,\FreqIdx\FrameIdx} =: \SpatialImage_{\MicIdx,\FreqIdx\FrameIdx} + \NoiseSig_{\MicIdx,\FreqIdx\FrameIdx}}
	\label{eq:mic_signal}
\end{equation}
where $\ATF_{\MicIdx,\FreqIdx}$, $\SrcSig_{\FreqIdx\FrameIdx}$, \WK{$\SpatialImage_{\MicIdx,\FreqIdx\FrameIdx}$} and $\NoiseSig_{\MicIdx,\FreqIdx\FrameIdx}$ \WK{denote} the \ac{ATF}, the anechoic source signal\WK{, the reverberant source signal at the microphone (spatial image),} and the $\MicIdx$-th additive noise signal comprising background noise and interfering signals, respectively. \WKtwo{Note that a two-microphone array is chosen for simplicity which does not limit generality of the proposed approach.} 
%The reverberant version of the anechoic source signal $\SrcSig_{\FreqIdx\FrameIdx}$ as observed at microphone $\MicIdx$ can be expressed by $\SpatialImage_{\MicIdx,\FreqIdx\FrameIdx} \coloneqq \ATF_{\MicIdx,\FreqIdx}\SrcSig_{\FreqIdx\FrameIdx}$ and is commonly termed spatial image. 
Knowledge of \acp{ATF} enables the construction of several powerful spatial filtering approaches including MPDR, MVDR and LCMV beamformers \cite{van_trees_optimum_2002,gannot_consolidated_2017}. However, as the estimation of \acp{ATF} is very difficult in practice, \acp{RTF} \cite{gannot_signal_2001,habets_application_2009} describing the relation between spatial images are employed for spatial filtering instead, as they can directly be estimated from the microphone signals \WK{during time intervals when noise and interference are negligibly small}. By taking (without loss of generality) the spatial image of the first microphone $\SpatialImage_{1,\FreqIdx\FrameIdx}$ as reference, the spatial image at the second microphone can be expressed via the \ac{RTF} $\RTF_{\FreqIdx}$ as $\SpatialImage_{2,\FreqIdx\FrameIdx} = \RTF_{\FreqIdx}\SpatialImage_{1,\FreqIdx\FrameIdx}$. This allows to describe both microphone signals in vector form by\vspace{-2pt}
\begin{equation}
	\begin{bmatrix}
		\MicSig_{1,\FreqIdx\FrameIdx}\\ \MicSig_{\MicIdxMax,\FreqIdx\FrameIdx}
	\end{bmatrix} = \begin{bmatrix}
	1\\ \RTF_\FreqIdx
\end{bmatrix}\SpatialImage_{1,\FreqIdx\FrameIdx} + \begin{bmatrix}
\NoiseSig_{1,\FreqIdx\FrameIdx}\\ \NoiseSig_{\MicIdxMax,\FreqIdx\FrameIdx}
\end{bmatrix}\in\mathbb{C}^\MicIdxMax.
	\label{eq:mic_signal_RTF}\vspace{-2pt}
\end{equation}
%While working well in noise-free scenarios, 
\WK{S}imple \ac{LS} estimators suffer from being biased when additive noise or interfering sources are \WK{present} \cite{shalvi_system_1996}. To address this problem, \ac{RTF} estimators that rely on knowledge of the statistical properties of additive noise \cite{markovich-golan_performance_2015} or employ specialized noise estimators \cite{cohen_relative_2004} have been proposed. Also deep learning-based methods have been applied for selecting useful frequency bins for \ac{RTF} estimation \cite{chazan_dnn-based_2018,malek_blockonline_2020}. When multiple point sources\eg speakers, are present, source separation methods like directionally constrained \ac{BSS} \cite{reindl_geometrically_2013,brendel_unified_2020} or simplex analysis \cite{laufer-goldshtein_source_2018} have been applied for \ac{RTF} estimation. However, under acoustically adverse conditions, \ac{RTF} estimation remains a challenging task and the resulting estimates may suffer from measurement errors which reduce the performance of subsequent spatial filtering tasks. 

Often, telecommunication devices are always used within the same acoustic environment\eg a hands-free communication system in a car or a teleconferencing toolkit built in a \WKtwo{personal computer}. Under this assumption, data collected under benign conditions can be used to train a model describing this acoustic environment. This idea has \WK{recently} been used \WK{to learn} dictionaries for modeling \acp{RTF} \cite{koldovsky_semi-blind_2013,koldovsky_dictionary-based_2021} or \acp{ATF} \cite{koren_supervised_2012,fozunbal_multi-channel_2008,haubner_online_2020}. While these dictionary-based methods represent linear models, it has been observed that \acp{RTF} are well-modeled by a nonlinear manifold \cite{laufer-goldshtein_study_2015}\ie the difference between \acp{RTF} is well-described by a nonlinear relation defined by a much lower number of degrees of freedom than \WK{the number of} \ac{RTF} filter taps. A manifold learning method for improving \ac{RTF} estimates based on diffusion maps has been proposed \cite{talmon_relative_2013,sofer_robust_2021} and a semi-supervised deep learning approach to infer \acp{RTF} from source positions has been developed in \cite{wang_semi-supervised_2018}. A \ac{VAE} has been leveraged for source localization from measured \acp{RTF} in \cite{bianco_semi-supervised_2020,bianco_semi-supervised_2021}.

In this paper, we propose a method of deep learning-based manifold learning for improving noisy \ac{RTF} estimates, \WK{which is not restricted to be linear such as dictionary-based methods and is based on a highly expressive model represented by a \ac{VAE}}. To this end, a \ac{VAE} unsupervisedly trained \WK{with data collected under benign acoustic conditions} to reconstruct \acp{RTF} of the considered enclosure is used to enhance inaccurate \acp{RTF} measured under acoustically adverse conditions. Furthermore, a hybrid of classic \ac{RTF} estimation and the trained \ac{VAE} is proposed. As an additional baseline assuming knowledge about the interfering signals, the trained \ac{VAE} is fine-tuned to take the noise conditions into account. Comprehensive experiments with real-world data featuring various acoustic conditions confirm the \WK{superiority} of the proposed method \WK{\WKtwo{relative to} several baselines}.
%
%The remainder of this paper is structured as follows: Sec.~\ref{sec:model}, Sec.~\ref{sec:setup}, Sec.~\ref{sec:results}, Sec.~\ref{sec:conclusion}\ToDo{not needed}
%
%%%%%%%%%%%%%%%%%%%%%%%%%%%%%%%%%%%%%%%%%%%%%%%%%%%%%%%%%
\section{Manifold Learning-based RTF Reconstruction}
\label{sec:model}
%%%%%%%%%%%%%%%%%%%%%%%%%%%%%%%%%%%%%%%%%%%%%%%%%%%%%%%%%
%
In Sec.~\ref{sec:VAE}, the \ac{VAE} concept is \WK{introduced for the \WKtwo{given} problem} and the training target for the \WK{considered} \ac{VAE} structure is introduced, \WK{which is used in} Sec.~\ref{sec:reconstruction} \WK{for} \ac{RTF} reconstruction.
%
%--------------------------------------------------------
\subsection{Variational Autoencoder}
\label{sec:VAE}
%--------------------------------------------------------
%
\acp{AE} \cite{hinton_reducing_2006} represent a powerful approach to unsupervised dimensionality reduction\ie to learn essential data representations and ignore insignificant data components. \WK{To this end}, a manifold representation of the input data is generated, accomplished by training a \ac{DNN} comprising an encoder and a decoder which are connected by a thin layer (the bottleneck) to reconstruct its input at its output. The output of the encoder, \WK{the embedding} $\EmbeddingVec\in\mathbb{R}^\BottleneckDim$, \WKtwo{constitutes} a low-dimensional representation of the input \WK{broadband \ac{RTF} \WKtwo{described} by real and imaginary parts}\vspace{-2pt}
\begin{gather}
	\RTFRI \coloneqq \big[\underbrace{\Real{\RTF_1},\dots,\Real{\RTF_\FreqIdxMax}}_{\RTFRI_{\mathrm{Re}}},\underbrace{\Imag{\RTF_1},\dots,\Imag{\RTF_\FreqIdxMax}}_{\RTFRI_{\mathrm{Im}}}\big]\transp\in \mathbb{R}^\InputSize,\raisetag{12pt}
	\label{eq:def_input}\vspace{-4pt}
\end{gather}
where $2\FreqIdxMax=\InputSize \gg \BottleneckDim$. To enforce structured embeddings\ie similar input data always correspond to similar embedding representations, \acp{VAE} \cite{kingma_auto-encoding_2013} have been proposed that employ stochastic encoders and decoders. Here, instead of trying to directly identify a (\WK{generally} intractable) generative model $\DecoderModel(\RTFRI)$ parameterized by the parameter vector $\DecoderParams$, the \ac{ELBO}
\begin{equation}
	\ELBO(\EncoderParams,\DecoderParams,\RTFRI)\coloneqq \Expect_{\EncoderModel(\EmbeddingVec\vert\RTFRI)}\log \DecoderModel(\RTFRI\vert\EmbeddingVec) - \KLD \left\lbrace \EncoderModel(\EmbeddingVec\vert\RTFRI)\Vert p(\EmbeddingVec) \right\rbrace,
	\label{eq:elbo}
\end{equation}
which bounds the log-likelihood of $\DecoderModel(\RTFRI)$ from below, is maximized \wrt $\EncoderParams$ and $\DecoderParams$ parameterizing the encoder and decoder, respectively. The second part of \eqref{eq:elbo} represents the \ac{KLD} between the prior $p(\EmbeddingVec)$ on the embedding $\EmbeddingVec$ and the encoder distribution $\EncoderModel(\EmbeddingVec\vert\RTFRI)$, which we choose to be normally distributed as
\begin{equation}
	p(\EmbeddingVec) \coloneqq \Gaussian(\mathbf{0}_\BottleneckDim,\mathbf{I}_{\BottleneckDim\times\BottleneckDim})\quad\text{and}\quad\EncoderModel(\EmbeddingVec\vert\RTFRI)\coloneqq \Gaussian(\MeanVec,\CovMat),
\end{equation}
respectively. The encoder distribution $\EncoderModel(\EmbeddingVec\vert\RTFRI)$, which represents a variational approximation of the true posterior of $\EmbeddingVec$ is parameterized by a mean vector $\MeanVec\in\mathbb{R}^\BottleneckDim$ and a diagonal covariance matrix $\CovMat\coloneqq\mathrm{diag}\{\VarianceVec\}\in\mathbb{R}^{\BottleneckDim\times\BottleneckDim}$. Both \WKtwo{parameter vectors, $\MeanVec$ and $\VarianceVec$,} are estimated from the input $\RTFRI$ with an encoder network (the application of the logarithm $\log\VarianceVec\in\mathbb{R}^\BottleneckDim$ is meant element-wise)
\begin{equation}
	\Encoder:\mathbb{R}^\InputSize \rightarrow \mathbb{R}^{2\BottleneckDim}\quad \mathrm{with}\quad (\MeanVec,\log \VarianceVec)\coloneqq\Encoder(\RTFRI).
\end{equation}
Hence, by maximizing \eqref{eq:elbo}, the \ac{KLD} term enforces the variational approximation $\EncoderModel(\EmbeddingVec\vert\RTFRI)$ to stay close to the prior $p(\EmbeddingVec)$.  The first term of \eqref{eq:elbo} is an expectation of the decoder distribution, which is chosen to be Gaussian\vspace{-7pt}
\begin{equation}
	\DecoderModel(\RTFRI\vert\EmbeddingVec)\coloneqq \Gaussian\Big(\RTFRIrec,\VarianceDec\, \mathbf{I}_{\InputSize\times\InputSize}\Big),
\end{equation}
where the mean vector $\RTFRIrec\in\mathbb{R}^{\InputSize}$ is estimated from the embedding $\EmbeddingVec$ by the decoder network\vspace{-2pt}
\begin{equation}
	\Decoder:\mathbb{R}^\BottleneckDim \rightarrow \mathbb{R}^{\InputSize}\quad \mathrm{with}\quad \RTFRIrec\coloneqq\Decoder(\EmbeddingVec)\vspace{-2pt}
\end{equation}
and $\VarianceDec$ is a hyperparameter (we use $\VarianceDec = 0.5$ in the following). When maximizing the \ac{ELBO}, the first term in \eqref{eq:elbo} enforces accurate reconstruction of the input at the output of the \ac{VAE}.

With the choices made for the encoder and decoder distribution, we employ the following cost function for training the \ac{VAE}\vspace{-2pt}
%\begin{align}
%	&\CostFunVAE(\EncoderParams,\DecoderParams,\RTFRI,\gamma)\coloneqq \gamma \frac{\sum_{\BatchIdx=1}^\BatchSize \Vert \RTFRI_\BatchIdx-\RTFRIrec_\BatchIdx\Vert_2^2}{\sum_{\BatchIdx=1}^\BatchSize\Vert \RTFRI_\BatchIdx\Vert_2^2}\cdots \label{eq:cost_function}\\
%	&\qquad\cdots\,-\frac{1-\gamma}{2\BatchSize\BottleneckDim}\sum_{\BatchIdx=1}^\BatchSize \left(\BottleneckDim +\log\det\CovMat_{\BatchIdx}-\Vert\MeanVec_{\BatchIdx}\Vert_2^2-\trace\,\CovMat_{\BatchIdx}\right).\nonumber
%\end{align}
\begin{align}
	&\CostFunVAE(\EncoderParams,\DecoderParams,\RTFRI,\gamma)\coloneqq \gamma\ \frac{\Avg_\BatchIdx \big\lbrace\Vert \RTFRI_\BatchIdx-\RTFRIrec_\BatchIdx\Vert_2^2\big\rbrace}{\Avg_\BatchIdx\left\lbrace\Vert \RTFRI_\BatchIdx\Vert_2^2\right\rbrace}\cdots \label{eq:cost_function}\\
	&\qquad\qquad\cdots\,-\frac{1-\gamma}{2\BottleneckDim}\ \Avg_\BatchIdx \left\lbrace\log\det\CovMat_{\BatchIdx}-\Vert\MeanVec_{\BatchIdx}\Vert_2^2-\trace\,\CovMat_{\BatchIdx}\right\rbrace.\nonumber\vspace{-4pt}
\end{align}
Here, a convex weight $\gamma\in[0,1]$ is introduced trading off the first part of $\CostFunVAE$ controlling the reconstruction performance against the second part of $\CostFunVAE$ resulting from the \ac{KLD} term in \eqref{eq:elbo} (cf. \cite{higgins_beta-vae_2017}). Note that $\gamma = 1$ will \WK{reduce} the model trained according to $\CostFunVAE$ to a deterministic \ac{AE}. To be independent of the input data's energy and vector lengths, we normalize the reconstruction term by the squared norms of the input vectors $\Vert \RTFRI_\BatchIdx\Vert_2^2$ and the \ac{KLD} term by the dimension of the embedding vectors $\BottleneckDim$. For robust gradient-descent-based optimization, averages $\Avg_\BatchIdx \{\cdot\} \coloneqq\frac{1}{\BatchSize}\sum_{\BatchIdx=1}^{\BatchSize}(\cdot)$ over batches with elements indexed by $\BatchIdx\in\{1,\dots,\BatchSize\}$ are incorporated into $\CostFunVAE$. 
%
%--------------------------------------------------------
\subsection{RTF Reconstruction}
\label{sec:reconstruction}
%--------------------------------------------------------
%
For \ac{RTF} estimation, we choose the well-known estimator \cite{shalvi_system_1996,gannot_signal_2001,cohen_relative_2004}\vspace{-3pt}
\begin{equation}
	\RTFest_\FreqIdx \coloneqq \frac{\Avg_\FrameIdx\{\AutoPSD\CrossPSD\} - \Avg_\FrameIdx\{\AutoPSD\}\Avg_\FrameIdx\{\CrossPSD\}}{\Avg_\FrameIdx\{\AutoPSD^2\} - \Avg_\FrameIdx\{\AutoPSD\}^2},
	\label{eq:RTF_estimator}\vspace{-3pt}
\end{equation}
which showed superior performance in our experiments \wrt simple \ac{LS}-based methods. Here, $\AutoPSD\coloneqq \vert \MicSig_{1,\FreqIdx\FrameIdx}\vert^2$ and $\CrossPSD\coloneqq\MicSig_{1,\FreqIdx\FrameIdx}\MicSig^\ast_{2,\FreqIdx\FrameIdx}$ denote instantaneous estimates of the auto and cross \ac{PSD} of the observed signals, respectively, and $\Avg_\FrameIdx$ a time average. Under acoustically adverse conditions, the estimated \acp{RTF} will be noisy and their usefulness for spatial filtering degrades. To enhance these \ac{RTF} estimates, we propose two approaches based on the trained \ac{VAE} model in the following. 

\noindent\textbf{Denoising \ac{VAE} (DN):} As the encoder network $\Encoder$ was trained to represent the most-essential components of an \ac{RTF}, measurement \WK{noise} that cannot be considered as a typical \ac{RTF} \WK{ingredient} will not be encoded by $\Encoder$. Hence, applying encoder and decoder successively to an estimated \ac{RTF} $\RTFRIest$ (defined analogously to \eqref{eq:def_input})\vspace{-3pt}
\begin{equation}
	\RTFRIrec = \Decoder\left(\MeanVec\right)\quad \mathrm{with}\quad (\MeanVec,\cdot)=\Encoder\big(\RTFRIest\big)
	\label{eq:denoising}\vspace{-3pt}
\end{equation}
will enhance the estimate $\RTFRIest$ by removing the measurement noise.

\noindent\textbf{\ac{VAE}-based \ac{LS} estimator (LS):} The \ac{RTF} reconstruction with DN only exploits a previously obtained \ac{RTF} estimate and \WK{neglects} the observed signals for its enhancement. In the following, we develop a reconstruction approach that takes both, an \ac{RTF} estimate and the observed signals, into account. By ignoring the additive noise in \eqref{eq:mic_signal_RTF}, the \ac{RTF} between the microphones could precisely be estimated by an \ac{LS} approach. To compensate for \WKtwo{neglecting} the noise, the reconstructed \ac{RTF} $\RTFRIrec$ is confined to the manifold learned by the \ac{VAE}
\begin{equation}
	\RTFRIrec= \left[\RTFRIrec\transp_{\mathrm{Re}}(\EmbeddingVec),\RTFRIrec\transp_{\mathrm{Im}}(\EmbeddingVec)\right]\transp\hspace{-3pt} = \Decoder(\EmbeddingVec)\ \ \ \ \text{with}\ \ \ \ \RTFRIrec_{\mathrm{Re}}(\EmbeddingVec),\RTFRIrec_{\mathrm{Im}}(\EmbeddingVec)\in\mathbb{R}^\FreqIdxMax\hspace{-5pt}.\nonumber
\end{equation}
The reconstructed \ac{RTF} $\RTFRIrec$ can now be determined by minimizing the \ac{LS} cost function
\begin{equation}
	\CostFunLS(\EmbeddingVec,\MicSigSet)\coloneqq\sum_{\FrameIdx=1}^\FrameIdxMax\left\Vert \MicSigVecBB_{1,\FrameIdx}\odot\left(\RTFRIrec_{\mathrm{Re}}(\EmbeddingVec)+\jmath\RTFRIrec_{\mathrm{Im}}(\EmbeddingVec)\right) -\MicSigVecBB_{2,\FrameIdx} \right\Vert_2^2
	\label{eq:LS_cost_function}
\end{equation}
subject to $\EmbeddingVec$. Here, we introduced the broadband observed signal vector
$\MicSigVecBB_{\MicIdx,\FrameIdx}\coloneqq [\MicSig_{\MicIdx,1\FrameIdx},\dots,\MicSig_{\MicIdx,\FreqIdxMax\FrameIdx}]\transp$, $\MicIdx\in\{1,2\}$, the set of observations
$\MicSigSet\coloneqq\big\lbrace\MicSigVecBB_{\MicIdx,\FrameIdx}\in\mathbb{C}^\FreqIdxMax\vert\MicIdx\in\{1,2\},\FrameIdx\in\{1,\dots,\FrameIdxMax\}\big\rbrace$ and the Hadamard product $\odot$. The \ac{LS} cost function \eqref{eq:LS_cost_function} is iteratively minimized by gradient descent ($\Decoder$ is differentiable)\vspace{-2pt}
\begin{equation}
	\MeanVec_{\mathrm{LS}} \leftarrow \MeanVec_{\mathrm{LS}} - \frac{\StepSizeLS}{\sum_{\FrameIdx=1}^\FrameIdxMax\Vert \MicSigVecBB_{1,\FrameIdx} \Vert_2^2} \frac{\mathrm{d}\CostFunLS(\EmbeddingVec,\MicSigSet)}{\mathrm{d}\EmbeddingVec} \bigg\vert_{\EmbeddingVec=\MeanVec_{\mathrm{LS}}},
	\label{eq:LS_update}
\end{equation}
where $\StepSizeLS\in\mathbb{R}_+$ denotes a step size. For stable convergence the gradient is normalized by the observed signal energy and the iterative process is initialized with the DN solution $(\MeanVec_{\mathrm{LS}},\cdot)=\Encoder\big(\RTFRIest\big)$. After termination of the iterative optimization \eqref{eq:LS_update}, the reconstructed \ac{RTF} is calculated by $\RTFRIrec_{\mathrm{LS}} = \Decoder\left(\MeanVec_{\mathrm{LS}}\right)$.
%
%%%%%%%%%%%%%%%%%%%%%%%%%%%%%%%%%%%%%%%%%%%%%%%%%%%%%%%%%
\section{Experimental Setup}
\label{sec:setup}
%%%%%%%%%%%%%%%%%%%%%%%%%%%%%%%%%%%%%%%%%%%%%%%%%%%%%%%%%
%
\begin{table}
	\centering
	\begin{tabular}{cl|lc}
		&\textbf{layer name} & \textbf{output shape} & \textbf{activation} \\
		\toprule
		\parbox[t]{2mm}{\multirow{5}{*}{\rotatebox[origin=c]{90}{$\Encoder$}}}&\texttt{input}: $\RTFRI_\BatchIdx$  & $\BatchSize\times \InputSize$ & - \\
		&\texttt{encoder1}               & $\BatchSize\times256$ & swish \\
		&\texttt{encoder2}			    & $\BatchSize\times128$ & swish \\
		&\texttt{encoder3}			    & $\BatchSize\times64$ & swish \\
		&\texttt{postPar}: $\MeanVec_\BatchIdx, \log\VarianceVec_\BatchIdx$	    & $\BatchSize\times\BottleneckDim\times2$ & linear \\
		\midrule
		&\texttt{sampling}: $\EmbeddingVec_\BatchIdx$   & $\BatchSize\times\BottleneckDim$ & - \\
		\midrule
		\parbox[t]{2mm}{\multirow{4}{*}{\rotatebox[origin=c]{90}{$\Decoder$}}}&\texttt{decoder1}			    & $\BatchSize\times64$ & swish \\
		&\texttt{decoder2}			    & $\BatchSize\times128$ & swish \\
		&\texttt{decoder3}			    & $\BatchSize\times256$ & swish \\
		&\texttt{output}: $\RTFRIrec_\BatchIdx$ & $\BatchSize\times \InputSize$ & linear \\
		\bottomrule
	\end{tabular}\vspace{-5pt}
	\caption{Proposed \ac{VAE} architecture, where $\BatchSize$ denotes the batch size and $\RTFRI_\BatchIdx,\RTFRIrec_\BatchIdx\in \mathbb{R}^\InputSize$, $\EmbeddingVec_\BatchIdx,\MeanVec_\BatchIdx,\VarianceVec_\BatchIdx\in\mathbb{R}^\BottleneckDim$.}
	\label{tab:VAE_architecture}\vspace{-8pt}
\end{table}
In the following section, the experimental setup, the proposed \ac{VAE} architecture, its training as well as the realization of the considered algorithmic variants is discussed.\\
\textbf{Datasets:} \WK{To} demonstrate the performance of the proposed method under real-world acoustic conditions, \acp{RTF} obtained from the MIRaGe dataset \cite{cmejla_mirage_2021} containing measurements from a varechoic lab of dimensions $6\,\mathrm{m}\times 6\,\mathrm{m}\times 2.4\,\mathrm{m}$ at a reverberation time of \mbox{$T_{60}\in\{0.1\,\mathrm{s},0.3\,\mathrm{s},0.6\,\mathrm{s}\}$} and speech signals from the ACE dataset \cite{eaton_estimation_2016} are used. To simulate noise and interferers present in real-world acoustic scenes, recorded noise signals including\eg meeting, factory, fan or vacuum cleaner noise, from the ACE \cite{eaton_estimation_2016} and NOISEX database \cite{varga_assessment_1993}, as well as own recordings, have been used.
The MIRaGe dataset contains recorded \ac{WGN} signals from a loudspeaker placed at grid positions of $2\,\mathrm{cm}$ spacing in $x$ and $y$ direction and $4\,\mathrm{cm}$ in $z$ direction within a cube of dimensions $46\,\mathrm{cm}\times 36\,\mathrm{cm}\times 32\,\mathrm{cm}$ at a height of $1.15\,\mathrm{m}$ of the grid center. The loudspeaker signal has been recorded by several spatially distributed microphone arrays from which we choose a microphone pair at $2\,\mathrm{m}$ distance from the grid center and same height with a spacing of $10\,\mathrm{cm}$. Additionally, several measurements at \acp{OOGP} at $1\,\mathrm{m}$ distance from the walls are available, which will be used for synthesizing additive noise signals. The dataset for training and evaluation of the proposed methods is created by estimating \acp{RTF} of $\InputSize\hspace{-1pt}=\hspace{-1pt}256$ taps length directly from the recorded \ac{WGN} signals\ie under optimal conditions. From all available $4104$ grid positions, we randomly select $\DataIdxMax\hspace{-1pt}=\hspace{-1pt}200$ for the test set and $100$ for the validation set. For data augmentation, the \acp{RTF} corresponding to the remaining $3804$ grid positions are repeated five times by adding \ac{WGN} with $1\,\%$ of the average \ac{RTF} variance yielding the training set.
\begin{figure}
	\tikzset{
	system/.style={draw,thick,rectangle},%
	branch dot/.style={draw,fill,circle,inner sep=1pt},%
	sum/.style={draw,thick,circle,inner sep=0pt},%
}
\newcommand{\suma}{\scriptsize$\bm{+}$}
\centering
\begin{tikzpicture}
	\node[system] at (-2.5,0.75)(expectation){Mean};
%	\node at (1.5,0.75)(meanRTF){Mean};
	
	\node[system] at (-2.5,-1.5)(estimator){Raw $\RTFRIest$};
	
	\node[system] at (0.5,0)(vae){VAE};
	
	\node[system, color=LMSgreen] at (-2.5,0)(RTFs){RTFs $\RTFRI$};
	\node[system] at (-3.5,-0.75)(noise){Noise};
	
	\node[branch dot, color=LMSred] at (1.5,0) (branchDot1) {};
	\node[system, color=LMSred] at (0.5,-0.75)(JLS){$\CostFunLS$};
	\node[system, color=LMSred] at (-0.5,-0.75)(micSigs){$\MicSigSet$};
	
	\node[system] at (0.5,-1.5)(vaeFT){VAE-FT};
	
	\node[system, color=LMSgreen] at (3.5,-0.6)(GT){GT};
	\node[system, color=LMSblue] at (3.5,0.6)(DN){DN};
	\node[system, color=LMSred] at (3.5,0)(LS){LS};
	
	\node[system, color=LMSblue] at (3.5,-1.2)(FT){FT};
	\node[system, color=LMSgreen] at (3.5,-1.8)(FTGT){FT-GT};

	\node[sum] at (2.5,0) (sum1) {\suma};
	\node[sum] at (-1.5,0) (sum2) {\suma};
	\node at (-1.35,0.2)(minus1){$\text{-}$};
	\node[sum] at (-1,-1.5) (sumFT1) {\suma};
	\node[sum] at (2,-1.5) (sumFT2) {\suma};
	\node at (-0.85,-1.3)(minus1){$\text{-}$};
	\node[sum] at (-2.5,-0.75) (sumNoise) {\suma};
	
	\draw[->, thick, color=LMSred] (branchDot1) |- (JLS);
	\draw[->, thick, color=LMSred] (micSigs) -- (JLS);
	\draw[->, thick, color=LMSred] (JLS) -- (vae);

	\node[branch dot] at (2,0.5) (branchDot2) {};
	\node[branch dot] at (-1,0.5) (branchDot3) {};
	\node[branch dot] at (0.5,0.5) (branchDot4) {};
	
	\draw[thick] (expectation) -| (branchDot4);
	\draw[thick] (expectation) -| (branchDot4);
	\draw[thick] (branchDot4) -- (branchDot2);
	\draw[thick] (branchDot4) -- (branchDot3);
	\draw[->, thick] (branchDot2) -| (sum1);
	\draw[->, thick] (branchDot3) -| (sum2);
	\draw[->, thick] (branchDot3) -| (sumFT1);
	\draw[->, thick] (branchDot2) -| (sumFT2);
	\draw[->, thick] (RTFs) -- (sumNoise);
	\draw[->, thick] (noise) -- (sumNoise);
	\draw[->, thick] (sumNoise) -- (estimator);
	\draw[->, thick, color=LMSblue] (estimator) -- (sum2);
	\draw[->, thick, color=LMSred] (sumNoise) -- (micSigs);
	
	\draw[->, thick, color=LMSblue] (estimator) -- (sumFT1);
	\draw[->, thick, color=LMSgreen] (RTFs) -- (sum2);
	\draw[->, thick, color=LMSgreen] (RTFs) -- (sumFT1);
	\draw[->, thick] (RTFs) -- (expectation);
	
	\draw[->, thick, color=LMSgreen] (sum2) -- (vae);
	\draw[->, thick, color=LMSblue] (-1.375,0.1) -- (0.075,0.1);
%	\draw[->, thick, color=LMSgreen] (-1.4,-0.1) -- (-0.4,-0.1);
	
	\draw[->, thick, color=LMSred] (vae) -- (sum1);
	\draw[->, thick, color=LMSblue] (0.92,0.1) -- (2.35,0.1);
	\draw[->, thick, color=LMSgreen] (0.92,-0.1) -- (2.35,-0.1);
	\draw[->, thick, color=LMSgreen] (sumFT1) -- (vaeFT);
	\draw[->, thick, color=LMSblue] (-0.875,-1.4) -- (-0.16,-1.4);
	\draw[->, thick, color=LMSgreen] (vaeFT) -- (sumFT2);
	\draw[->, thick, color=LMSblue] (1.15,-1.4) -- (1.85,-1.4);
	
	\draw[->, thick, color=LMSgreen] (sum1) -- (GT);
	\draw[->, thick, color=LMSblue] (sum1) -- (DN);
	\draw[->, thick, color=LMSred] (sum1) -- (LS);
	
	\draw[->, thick, color=LMSblue] (sumFT2) -- (FT);
	\draw[->, thick, color=LMSgreen] (sumFT2) -- (FTGT);
\end{tikzpicture}\vspace{-10pt}
	\caption{Overview of the considered algorithmic variants.}
	\label{fig:block_diagram}\vspace{-10pt}
\end{figure}
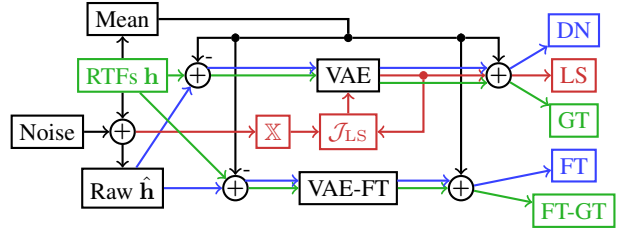

\noindent\textbf{Microphone signals:} To generate reverberant observations, we use a microphone placed closely to the loudspeaker \WK{representing the source} during the measurements and estimate the \acp{ATF} to the considered microphone pair. The microphone signals are obtained by filtering anechoic speech signals of $10\,\mathrm{s}$ duration from the ACE dataset with the previously estimated \acp{ATF} and adding noise signals at a desired \ac{SNR}. Various types of additive noise are considered: \ac{WGN}, interfering speech or recorded noise signals at a specific \ac{OOGP}\ie a \ac{PS}, and noise signals at multiple available \acp{OOGP}\ie approximating a diffuse noise source. The interfering speech signals (not used for the desired source) are taken from the ACE database \cite{eaton_estimation_2016}. 
%The recorded noise signals include\eg meeting, factory, fan or vacuum cleaner noise.
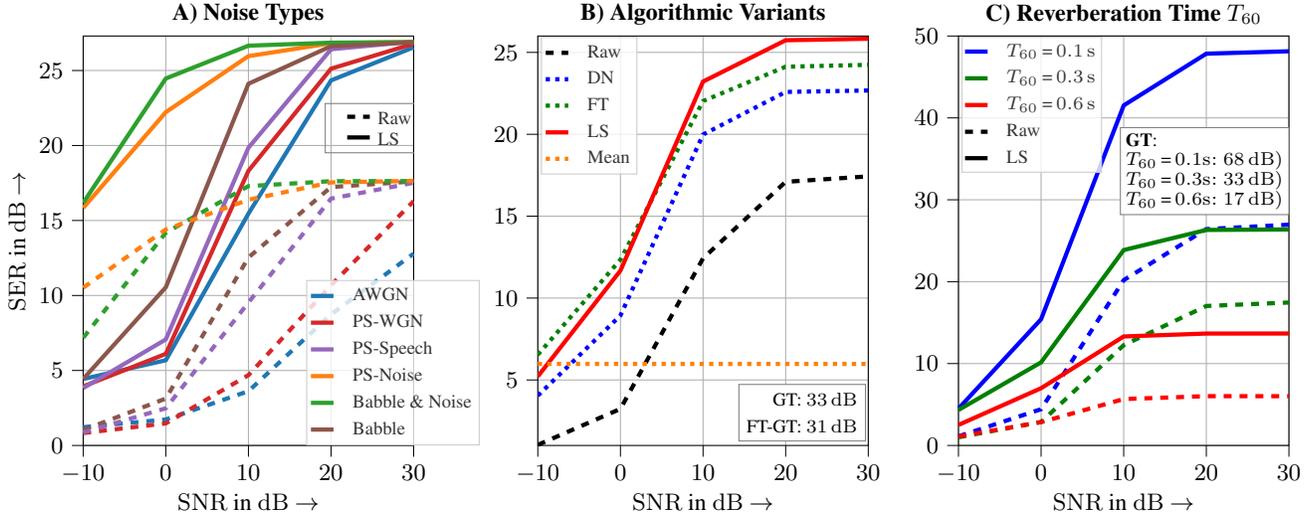
\begin{figure*}
	\centering
	% This file was created with tikzplotlib v0.9.12.
\begin{tikzpicture}

\definecolor{color0}{rgb}{0.12156862745098,0.466666666666667,0.705882352941177}
\definecolor{color1}{rgb}{0.83921568627451,0.152941176470588,0.156862745098039}
\definecolor{color2}{rgb}{0.549019607843137,0.337254901960784,0.294117647058824}
\definecolor{color3}{rgb}{0.580392156862745,0.403921568627451,0.741176470588235}
\definecolor{color4}{rgb}{0.172549019607843,0.627450980392157,0.172549019607843}
\definecolor{color5}{rgb}{1,0.498039215686275,0.0549019607843137}

\begin{axis}[
width = 170,
height = 200,
legend cell align={left},
legend style={
  fill opacity=0.8,
  draw opacity=1,
  text opacity=1,
  at={(1.2,0.0)},
  anchor=south east,
  draw=white!80!black,
  font = \scriptsize,
  inner sep = 0.75pt
},
legend columns = 1,
legend image post style={scale=0.5},
tick align=outside,
tick pos=left,
title style={font=\bfseries, yshift = -5pt},
title={A) Noise Types},
x grid style={white!69.0196078431373!black},
xlabel = {$\mathrm{SNR}$ in $\mathrm{dB}$ $\rightarrow$},
xmajorgrids,
xmin=-10, xmax=30,
xtick style={color=black},
y grid style={white!69.0196078431373!black},
ylabel style = {yshift = -10pt},
ylabel={$\mathrm{SER}$ in $\mathrm{dB}$ $\rightarrow$},
ymajorgrids,
ymin=0, ymax=27.3,
ytick style={color=black}
]

\addplot [ultra thick, dashed, color0, forget plot]
table {%
-10 1.21820553329779
0 1.73301445450813
10 3.618447724982
20 8.74114053934738
30 12.7887329807969
};
%\addplot [ultra thick, color0, dashed, forget plot]
%table {%
%-10 4.16722497151008
%0 5.39772127671827
%10 13.3783011352454
%20 22.2592526150177
%30 25.0148775398476
%};
\addplot [ultra thick, color0]
table {%
-10 4.43050953764369
0 5.68111377855444
10 15.4568709737028
20 24.3399569980999
30 26.5467440206969
};
\addlegendentry{AWGN}
\addplot [ultra thick, dashed, color1, forget plot]
table {%
-10 0.827986117042548
0 1.4721366201194
10 4.68747358642143
20 10.6799980290688
30 16.2907460948663
};
%\addplot [ultra thick, color1, dashed, forget plot]
%table {%
%-10 3.89361652531616
%0 5.79469424635618
%10 15.5815727752027
%20 22.9871990137427
%30 25.1107066539117
%};
\addplot [ultra thick, color1]
table {%
-10 3.93625838151354
0 6.10682173658974
10 18.3129583205225
20 25.1216299018483
30 26.7856826956092
};
\addlegendentry{PS-WGN}
\addplot [ultra thick, dashed, color2, forget plot]
table {%
-10 1.02753692794876
0 3.12777605163799
10 12.5316931286196
20 17.22898529976
30 17.6054514590125
};
%\addplot [ultra thick, color2, dashed, forget plot]
%table {%
%-10 4.32728454812054
%0 9.23210070339264
%10 22.2972227625986
%20 24.9956770793727
%30 25.2771076416598
%};
\addplot [ultra thick, dashed, color3, forget plot]
table {%
-10 0.867018083860469
0 2.47758495778169
10 9.53051500884373
20 16.4584950433146
30 17.5144517783395
};
%\addplot [ultra thick, color3, dashed, forget plot]
%table {%
%-10 3.87692675083118
%0 6.59687890742579
%10 17.3566594869918
%20 24.6462949114482
%30 25.3323385867979
%};
\addplot [ultra thick, color3]
table {%
-10 3.82328267734095
0 7.06957972778689
10 19.8503827693792
20 26.4178238126526
30 26.9306753445511
};
\addlegendentry{PS-Speech}
\addplot [ultra thick, dashed, color4, forget plot]
table {%
-10 7.17943690614085
0 14.1819891501594
10 17.292202431385
20 17.6216241548173
30 17.629807046567
};
\addplot [ultra thick, color5]
table {%
	-10 15.8355561733654
	0 22.220965894207
	10 25.9585668627168
	20 26.8176382120917
	30 26.9012866973388
};
\addlegendentry{PS-Noise}
%\addplot [ultra thick, color4, dashed, forget plot]
%table {%
%-10 16.8041139733973
%0 24.1172047901397
%10 25.240982455935
%20 25.2796251052346
%30 25.3012877531546
%};
\addplot [ultra thick, color4]
table {%
-10 16.1765891316455
0 24.4661453831263
10 26.6548799616458
20 26.8561634428766
30 26.8909922539486
};
\addplot [ultra thick, color2]
table {%
	-10 4.43481968493019
	0 10.5440925686162
	10 24.1138421753364
	20 26.6213050869175
	30 26.8781649710737
};
\addlegendentry{Babble \& Noise}
\addlegendentry{Babble}
\addplot [ultra thick, dashed, color5, forget plot]
table {%
-10 10.5173736533216
0 14.4026444863041
10 16.3816890199453
20 17.5401637077442
30 17.6409071564977
};
%\addplot [ultra thick, color5, dashed, forget plot]
%table {%
%-10 17.8768067266276
%0 22.816373632534
%10 24.5231729430808
%20 25.2546583544701
%30 25.3154330697674
%};
%\addplot [line width=0pt, color4]
%table {%
%0 0
%0 0
%};
%\addlegendentry{RTF estimator}
%\addplot [thick, black]
%table {%
%0 0
%0 0
%};
%\addlegendentry{Raw}
%\addplot [thick, black, dash pattern=on 1pt off 3pt on 3pt off 3pt]
%table {%
%0 0
%0 0
%};
%\addlegendentry{VAE-D}
%\addplot [thick, black, dotted]
%table {%
%0 0
%0 0
%};
%\addlegendentry{LS}

\addlegendimage{ultra thick, solid, color=black}
\label{LS}
%\addlegendentry{LS}
\addlegendimage{ultra thick, dashed, color=black}
\label{raw}
%\addlegendentry{Raw}

\node [draw = gray, fill opacity=0.8,draw opacity=1,text opacity=1] at (rel axis cs: 0.875,0.775) {\shortstack[l]{
		\ref{raw} \scriptsize Raw \\
		\ref{LS} \scriptsize LS}};

\end{axis}

\end{tikzpicture}% This file was created with tikzplotlib v0.9.12.
\begin{tikzpicture}

\definecolor{color0}{rgb}{0.12156862745098,0.466666666666667,0.705882352941177}
\definecolor{color1}{rgb}{0.647058823529412,0.164705882352941,0.164705882352941}

\begin{axis}[
width = 170,
height = 200,
legend cell align={left},
legend style={
  fill opacity=0.8,
  draw opacity=1,
  text opacity=1,
  at={(0.0095,0.998)},
  anchor=north west,
  draw=white!80!black,
  font = \scriptsize,
  inner sep = 0.75pt
},
legend image post style={scale=0.5},
tick align=outside,
tick pos=left,
title style={font=\bfseries, yshift = -5pt},
title={B) Algorithmic Variants},
x grid style={white!69.0196078431373!black},
xlabel = {$\mathrm{SNR}$ in $\mathrm{dB}$ $\rightarrow$},
xmajorgrids,
xmin=-10, xmax=30,
xtick style={color=black},
y grid style={white!69.0196078431373!black},
ymajorgrids,
ymin=1, ymax=26,
ytick style={color=black}
]

\node[fill = white, draw = gray, align = right] at (axis cs: 22,3) {\scriptsize GT: $33\,\mathrm{dB}$\\ \scriptsize FT-GT: $31\,\mathrm{dB}$};

\addplot [ultra thick, black, dashed]
table {%
-10 1.03472426160823
0 3.23293830013513
10 12.4308282370175
20 17.0968567382578
30 17.4301822875043
};
\addlegendentry{Raw}
\addplot [ultra thick, blue, dotted]
table {%
-10 4.04219551357985
0 8.9334628899717
10 19.9837045569827
20 22.5851788243182
30 22.6736069494031
};
\addlegendentry{DN}
\addplot [ultra thick, green!50.1960784313725!black, dotted]
table {%
-10 6.53959669197355
0 12.3402287128822
10 22.0283860708149
20 24.1249624172937
30 24.2407054699615
};
\addlegendentry{FT}
\addplot [ultra thick, red]
table {%
-10 5.21579861324863
0 11.6726355520991
10 23.2186573191132
20 25.7417528806825
30 25.8336018836879
};
\addlegendentry{LS}
\addplot [ultra thick, orange, dotted]
table {%
-10 5.97837250456822
0 5.97837250456822
10 5.97837250456822
20 5.97837250456822
30 5.97837250456822
};
\addlegendentry{Mean}
\addplot [ultra thick, black]
table {%
-10 33.3102604382936
0 33.3102604382936
10 33.3102604382936
20 33.3102604382936
30 33.3102604382936
};
%\addlegendentry{VAE-D-GT}
\addplot [ultra thick, black, dashed]
table {%
-10 31.1479351630214
0 31.1479351630214
10 31.1479351630214
20 31.1479351630214
30 31.1479351630214
};
%\addlegendentry{VAE-FT-GT}
\end{axis}

\end{tikzpicture}% This file was created by matlab2tikz.
\definecolor{mycolor1}{rgb}{1.00000,0.00000,1.00000}%
\definecolor{mycolor2}{rgb}{0.00000,1.00000,1.00000}%
\definecolor{color0}{rgb}{0.12156862745098,0.466666666666667,0.705882352941177}
\definecolor{color1}{rgb}{0.647058823529412,0.164705882352941,0.164705882352941}
\begin{tikzpicture}
	
\begin{axis}[%
width = 170,
height = 200,
legend cell align={left},
legend style={
	fill opacity=0.8,
	draw opacity=1,
	text opacity=1,
	at={(0.0095,0.998)},
	anchor=north west,
	draw=white!80!black,
	font = \scriptsize,
	inner sep = 0.25pt
},
legend entries={$T_{60}\,\text{=}\, 0.1\,\mathrm{s}$,
	$T_{60}\,\text{=}\,0.3\,\mathrm{s}$,
	$T_{60}\,\text{=}\,0.6\,\mathrm{s}$,
	Raw,
%	VAE-D,
	LS},
legend image post style={scale=0.5},
tick align=outside,
tick pos=left,
title style={font=\bfseries, yshift = -5pt},
title={C) Reverberation Time $T_{60}$},
x grid style={white!69.0196078431373!black},
xlabel = {$\mathrm{SNR}$ in $\mathrm{dB}$ $\rightarrow$},
xmajorgrids,
xmin=-10, xmax=30,
xtick style={color=black},
y grid style={white!69.0196078431373!black},
ylabel style = {yshift = -10pt},
ymajorgrids,
ymin=0, ymax=50,
ytick style={color=black}
]
\addlegendimage{ultra thick, color=blue}
\addlegendimage{ultra thick, color=green!50.1960784313725!black}
\addlegendimage{ultra thick, color=red}
\addlegendimage{ultra thick, dashed, color=black}
\addlegendimage{ultra thick, solid, color=black}

\node[fill = white, draw = gray, align = left, inner sep=2, text width=2.1cm] at (axis cs: 19.8,33.5) {\scriptsize \textbf{GT}:\\ \vspace{-3pt}
\scriptsize$T_{60}\,\text{=}\,0.1\mathrm{s}\hspace{-2pt}: 68\,\mathrm{dB})$\vspace{-3pt}
$T_{60}\,\text{=}\,0.3\mathrm{s}\hspace{-2pt}: 33\,\mathrm{dB})$\vspace{-3pt}
$T_{60}\,\text{=}\,0.6\mathrm{s}\hspace{-2pt}: 17\,\mathrm{dB})$};

\addplot [ultra thick,color=blue, dashed]
table{%
-10 1.11289155265625
0 4.42753479502479
10 20.1917833892636
20 26.4202679813493
30 26.9601240797446
};
%\addlegendentry{Raw RTF}

%\addplot [ultra thick,color=blue, dashed]
%table{%
%-10 4.29946777968984
%0 13.6358098705884
%10 39.2303881159857
%20 46.2759791409255
%30 46.5814370373099
%};
%\addlegendentry{VAE-D}

\addplot [ultra thick,color=blue]
table{%
-10 4.49669225817962
0 15.4133955617886
10 41.523195942109
20 47.8424389757718
30 48.1351937269704
};
%\addlegendentry{VAE-LS}

%\addplot [color=black, dashed]
%  table[row sep=crcr]{%
%-10	5.53\\
%0	5.53\\
%10	5.53\\
%20	5.53\\
%30	5.53\\
%};
%\addlegendentry{Mean RTF}

%\addplot [color=mycolor2, dashed]
%  table[row sep=crcr]{%
%-10	68.1\\
%0	68.1\\
%10	68.1\\
%20	68.1\\
%30	68.1\\
%};
%\addlegendentry{VAE-D-GT}

\addplot [ultra thick,color=green!50.1960784313725!black, dashed, forget plot]
table {%
	-10 0.997841550603608
	0 2.86941324271909
	10 12.2483751911245
	20 17.0303331680468
	30 17.4544890378078
};
%\addplot [ultra thick,color=blue, forget plot]
%table {%
%	-10 4.33344170581255
%	0 9.21859051213607
%	10 21.811137266092
%	20 24.7183852506639
%	30 24.6915740700056
%};
\addplot [ultra thick,color=green!50.1960784313725!black, forget plot]
table {%
	-10 4.33197223753026
	0 10.1496549304427
	10 23.8614007442836
	20 26.3189900548797
	30 26.3804887886607
};
\addplot [ultra thick,color=red, dashed, forget plot]
table {%
	-10 1.10112372618193
	0 2.84697699932391
	10 5.66646817702376
	20 6.02086813172979
	30 6.01508612525661
};
%\addplot [ultra thick,color=blue, dotted, forget plot]
%table {%
%	-10 2.43047184916602
%	0 6.23117243437069
%	10 11.6842318231574
%	20 12.004278855725
%	30 12.0142195484141
%};
\addplot [ultra thick,color=red, forget plot]
table {%
	-10 2.50542322539061
	0 6.99116459749811
	10 13.3107963710176
	20 13.6657025167608
	30 13.6704513334879
};
\end{axis}
\end{tikzpicture}%
\vspace{-10pt}
	\caption{\textbf{A)} Influence of various noise types on \ac{RTF} estimation and the proposed LS approach. \textbf{B)} Comparison of all considered algorithmic variants. \textbf{C)} Influence of $T_{60}$ on \ac{RTF} estimation and the proposed LS approach.}
	\label{fig:results}\vspace{-10pt}
\end{figure*}

\noindent\textbf{\ac{VAE} network and training:} The proposed \ac{VAE} architecture summarized in Tab.~\ref{tab:VAE_architecture} takes minibatches $[\RTFRI_1,\dots,\RTFRI_\BatchSize]\transp\in\mathbb{R}^{\BatchSize\times\InputSize}$ of batch size $\BatchSize\hspace{-1pt}=\hspace{-1pt}128$ as input with elements $\RTFRI_\BatchIdx\in\mathbb{R}^\InputSize$ randomly chosen from the training set. The inputs are processed with three fully-connected layers and swish activation \cite{ramachandran_searching_2017} followed by a linear layer $\texttt{postPar}$ for the estimation of the parameters of the encoder distribution $\EncoderModel(\EmbeddingVec\vert\RTFRI)$. To enable backpropagation through the network, $\texttt{sampling}$ realizes the `reparameterization trick' \cite{kingma_auto-encoding_2013}\vspace{-2pt}
\begin{equation}
	\EmbeddingVec_\BatchIdx = \MeanVec_\BatchIdx + \exp\left(\frac{\log\VarianceVec_\BatchIdx}{2}\right)\odot\mathbf{e}_\BatchIdx\quad\text{with}\quad\mathbf{e}_\BatchIdx\sim\mathcal{N}(\mathbf{0}_\BottleneckDim,\mathbf{I}_{\BottleneckDim\times\BottleneckDim}).
	\label{eq:reparameterization_trick}\vspace{-2pt}
\end{equation}
\WKtwo{The bottleneck dimension was empirically chosen as $\BottleneckDim=5$.} Again, $\log$ and $\exp$ in \eqref{eq:reparameterization_trick} denote element-wise operations. The decoder is symmetric to the encoder structure resulting in $\approx 2.15\cdot10^5$ trainable parameters in total. To avoid learning of \ac{RTF} components common to all \acp{RTF} in the dataset, their mean is subtracted, the \ac{VAE} is trained on the residual and the mean \ac{RTF} is added to the output of the \ac{VAE} again for reconstruction. The network is trained by minimizing \eqref{eq:cost_function} with $\gamma= 0.95$ by ADAM \cite{kingma_adam_2017} with an initial learning rate of $10^{\text{-}3}$ which is reduced by a factor of five to avoid getting stuck with the training process if the validation loss did not improve by at least $10^{\text{-}3}$ within the last five epochs. To avoid overfitting, early stopping is employed and the network parameters of the epoch with lowest validation loss are restored if the validation loss did not improve by at least $10^{\text{-}3}$ within the last ten epochs.

In the remainder of the paper, we discuss and experimentally evaluate the following algorithmic variants \WK{illustrated in Fig.~\ref{fig:block_diagram}}:

\noindent\textbf{Raw \ac{RTF} estimation (Raw):} All algorithmic variants are based on \acp{RTF} estimated by \eqref{eq:RTF_estimator}. Hence, these raw \ac{RTF} estimates represent the first baseline for the experimental comparison.\\
\noindent\textbf{Mean of dataset (Mean):} The \ac{VAE} learns an \ac{RTF} representation as a refinement of the \WK{average of all \acp{RTF} in the training set. Hence, this mean \ac{RTF} represents} another natural baseline.\\
\noindent\textbf{Denoising \ac{VAE} (DN):} The estimated \acp{RTF} are denoised by \eqref{eq:denoising}.\\
\noindent\textbf{\ac{VAE}-based \ac{LS} estimator (LS):}  The \ac{RTF} is estimated by minimizing $\CostFunLS$ by \eqref{eq:LS_update} with $\alpha\hspace{-1pt}=\hspace{-1pt}2$. Here, only 20 iterations are executed to avoid overfitting to the interfering signals.\\
\noindent\textbf{Denoising with fine-tuned \ac{VAE} (FT):} Assuming that pairs of estimated noisy and clean \acp{RTF} are available, the \ac{VAE} can be trained to map noisy \acp{RTF} to clean ones instead of reconstructing the input \ac{RTF} at the output. In this way the typical deterioration of estimated \acp{RTF} is taken into account. To this end, we fine-tune the \ac{VAE} trained on clean \acp{RTF} by continuing training with a set of noisy and clean \acp{RTF} for additional $15$ epochs with the Adam optimizer with a learning rate of $10^{\text{-}4}$. However, such a training set is rarely given in practice and, hence, FT should be considered as an oracle baseline.\\
\noindent\textbf{Decoding of ground truth \ac{RTF} (GT):} \WK{As} an upper bound on the expected performance, the \ac{GT} \ac{RTF} is reconstructed by the \ac{VAE} similar to \eqref{eq:denoising}: $\RTFRIrec = \Decoder\left(\MeanVec\right)$ with $(\MeanVec,\cdot)=\Encoder\big(\RTFRI\big)$.
In this way, the modeling capability of the \ac{VAE} is evaluated as an upper bound for comparison to the proposed methods. Similarly, we denote the reconstruction of the \ac{GT} \ac{RTF} with FT, as \textbf{FT-GT}.

The performance of all investigated \ac{RTF} estimation methods is measured by the \ac{SER} (larger values correspond to better performance)\vspace{-2pt}
\begin{equation}
	\SER \coloneqq \frac{10}{\DataIdxMax}\sum_{\DataIdx=1}^\DataIdxMax \log_{10} \frac{\Norm{\RTFRI_\DataIdx}^2}{\Vert\RTFRI_\DataIdx - \RTFRIrec_\DataIdx\Vert_2^2},\vspace{-2pt}
\end{equation}
where the samples of the test set are indexed by $\DataIdx\in\{1,\dots,\DataIdxMax\}$.
%
%%%%%%%%%%%%%%%%%%%%%%%%%%%%%%%%%%%%%%%%%%%%%%%%%%%%%%%%%
\section{Results}
\label{sec:results}
%%%%%%%%%%%%%%%%%%%%%%%%%%%%%%%%%%%%%%%%%%%%%%%%%%%%%%%%%
%
The experimental results corresponding to the methodology described in Sec.~\ref{sec:setup} are shown in Fig.~\ref{fig:results} and are discussed below:

\noindent\textbf{A) Noise Types}: On the left of Fig.~\ref{fig:results}, LS and Raw are evaluated for various noise types at $T_{60} = 0.3\,\mathrm{s}$, where it can be seen that LS achieves an improvement over Raw up to about $10\,\mathrm{dB}$ for some scenarios. \ac{WGN} added to the microphone signals (AWGN) caused the worst performance in these experiments. As a second kind of noise, randomly placed \acp{PS} for the evaluation of each of the $\DataIdxMax=200$ test \acp{RTF}, which emitted either \ac{WGN}, speech or a recorded noise signal, simulate scenarios with an interfering source\ie an undesired speaker or a compact noise source in the background. Here, LS achieves similar results for \ac{WGN} and speech, which are also close to the results for AWGN. An interfering noise \ac{PS} was less detrimental in our experiments. To simulate ambient noise, we place in a third set of experiments several \acp{PS} at \acp{OOGP} emitting speech or recorded noise signals. We denote these experimental conditions by `Babble' if only speech signals are used and `Babble \& Noise' if a set containing speech and recorded noise signals is distributed over the \acp{OOGP}. Here, the `Babble \& Noise' performs similar to a \ac{PS} noise interferer and `Babble' shows slightly better but similar performance than a \ac{PS} speech interferer.

\noindent\textbf{B) Algorithmic Variants}: In the middle of Fig.~\ref{fig:results}, the different algorithmic variants described in Sec.~\ref{sec:setup} are compared experimentally for $T_{60} = 0.3\,\mathrm{s}$ and additive `Babble' noise. The \ac{SER} of the raw \ac{RTF} estimate is increased by DN, FT and LS, where DN yields the lowest improvements. While slightly worse than FT for very low \acp{SNR}, LS yields the best results of the considered enhancement approaches above $\mathrm{SNR}=10\,\mathrm{dB}$. However, it should be noted that while FT assumes knowledge about estimation errors as outlined in Sec.~\ref{sec:setup}, LS does not need such prior knowledge. DN, FT and LS show better performance than the trivial baseline represented by the mean of the \ac{RTF} data set (Mean) whereas Raw becomes better than Mean for \acp{SNR} above $0\,\mathrm{dB}$. The \ac{SER} of the reconstructed \ac{GT} \acp{RTF} is slightly lower for FT ($31\,\mathrm{dB}$) than without fine tuning ($33\,\mathrm{dB}$), which is to be expected as FT is optimized for noisy input and clean output \acp{RTF} and not for perfect reconstruction of \acp{RTF}. 

\noindent\textbf{C) Reverberation Time $T_{60}$}: On the right of Fig.~\ref{fig:results} the achieved \ac{SER} of LS and Raw is shown for additive `Babble' noise and varying $T_{60}$. LS significantly improves the \ac{SER} of Raw for all $T_{60}$, where the largest improvements are obtained for $T_{60} = 0.1\,\mathrm{s}$. The overall \ac{RTF} estimation performance as well as the GT decreases with increasing $T_{60}$, which is to be expected as the more and more complex \acp{RTF} are modeled with a filter of same length $\InputSize$. 
%
%%%%%%%%%%%%%%%%%%%%%%%%%%%%%%%%%%%%%%%%%%%%%%%%%%%%%%%%%
\section{Conclusion}
\label{sec:conclusion}
%%%%%%%%%%%%%%%%%%%%%%%%%%%%%%%%%%%%%%%%%%%%%%%%%%%%%%%%%
%
In this contribution, we propose a \ac{VAE}-based manifold model for \acp{RTF} and leverage it for enhancing \ac{RTF} estimates. We show that the proposed \ac{LS}-based \ac{RTF} estimator regularized by the trained \ac{VAE} increases the \ac{RTF} quality relative to \ac{VAE}-based denoising of the \ac{RTF} estimates. All experiments are conducted with measured data which emphasizes the real-world applicability of the proposed method. As next steps, we will combine and evaluate the proposed method with spatial filtering algorithms. A further improvement over the proposed method is expected by employing complex-valued networks and estimators for the statistics of additive noise.
% -------------------------------------------------------------------------
\bibliographystyle{IEEEbib}
\bibliography{refs}

\end{document}